\documentclass[aps,prf,superscriptaddress,longbibliography]{revtex4-2}

\usepackage{todonotes}
\usepackage{amsmath,amssymb,graphicx,float,epstopdf}
\usepackage{dcolumn}% Align table columns on decimal point
\usepackage[table]{xcolor}% Allows greying of cells in table
\usepackage{bm}% bold math
\usepackage{color}
\graphicspath{{figures}}

\usepackage{graphicx}  % Allows image embedding
\usepackage{float}
\usepackage[format=hang]{caption}

\usepackage{tikz}
\usetikzlibrary{arrows.meta,bending,positioning,intersections}
\usepackage{pgfplots} %% XY-pic package to draw
\usepackage[all]{xy} %gets rid of bottom navigation bars
\usepackage{wrapfig} %allows text to be wrapped around a figure

\definecolor{ForestGreen}{rgb}{0.2, 0.5, 0.2}

\usepackage[normalem]{ulem} % To strikeout text so that editing is easier

\begin{document}

\title{Brain pulsations enhance cerebrospinal fluid flow in  perivascular spaces
}

\author{Gregory Holba}
\email[]{greg.holba@open.ac.uk}
\affiliation{School of Mathematics and Statistics, The Open University, Milton Keynes MK7 6AA, United Kingdom}
\author{James P.~Hague}
\email[]{jim.hague@open.ac.uk}
\affiliation{School of Physical Sciences, The Open University, Milton Keynes MK7 6AA, United Kingdom}
\author{Nigel Hoggard}
\email[]{nigel.hoggard@nhs.net}
\affiliation{School of Medicine and Population Health, University of Sheffield, Sheffield S10 2TN, United Kingdom}
\affiliation{Sheffield NIHR Biomedical Research Centre, Royal Hallamshire Hospital, Sheffield S10 2JF, United Kingdom}
\author{Marc Pradas}
\email[]{marc.pradas@open.ac.uk}
\affiliation{School of Mathematics and Statistics, The Open University, Milton Keynes MK7 6AA, United Kingdom}

\date{\today}

\begin{abstract}
\noindent A novel approach is adopted to model cerebrospinal fluid (CSF) flow in human perivascular spaces (PVSs) surrounding brain-penetrating arteries. It is proposed that the outer PVS boundary oscillates due to brain pulsations and the arterial wall motion is driven by a blood pressure wave. 
Lubrication theory is employed to derive a mathematical model for the CSF flow, which is then solved numerically. A parametric analysis is undertaken to investigate the effect of the brain pulsations, which shows that pulsations magnify the net axial CSF flows created by the arterial wall motion. The findings suggest that net axial CSF flows are almost entirely positive (deeper into the brain), with arterial wall motion highly dependent on PVS-penetrating artery configurations. Given the glymphatic hypothesis, the findings support the clinical practice of treating dilated PVS as indicators of an increased likelihood of neurodegenerative conditions, such as dementia. 
\end{abstract}
% insert suggested PACS numbers in braces on next line
%\keywords{brain pulsations, cerebrospinal spinal fluid, CSF, fluid dynamics, perivascular pumpings, lubrication theory, parametric study, human brain studies }
\maketitle %\maketitle must follow title, authors, abstract and \pacs

\section{Introduction}

Cerebrospinal fluid (CSF) flows within and around the brain, providing protection, delivering nutrients and carrying away waste products that have been implicated in neurodegenerative conditions~\cite{Spector15, Wichman22}. Without effective brain waste clearance neurodegenerative conditions can develop, such as Alzheimer’s and Parkinson’s diseases, with dementia currently affecting over 55 million people worldwide~\cite{THELANCETNEUROLOGY2023643}. Given the increasing prevalence of such conditions~\cite{Collaborators22}, understanding the causes of these diseases is becoming imperative.

The flow of CSF is of particular interest within the narrow perivascular spaces (PVSs) surrounding brain penetrating arteries, as these are thought to be the entry points of the glymphatic system, the hypothesised waste clearance system of the brain~\cite{Iliff12, Nedergaard16, Kylkilahti21, Bohr22}. It is argued that without sufficient CSF flow in the PVS, waste clearance is restricted and unwanted proteins can accumulate and lead to the pathologies mentioned earlier. 

Our review of experimental research examining CSF in human brains was unable to identify any reports of direct measurements of CSF flows in brain penetrating arterial PVSs. This is because of the limited resolution of technologies available to map such PVS anatomy and accurately measure \emph{in-vivo} CSF flows~\cite{Zhao20}, alongside ethical research considerations. The vast majority of related studies have employed functional magnetic resonance imaging (fMRI) to examine aspects of CSF flow and include phase-contrast~\cite{Odeen11, Shi20}, contrast-enhanced~\cite{Yamamoto24}, 4D flow~\cite{Vikner24} and cine phase contrast~\cite{Bhadelia95, Linninger09} methods. 

A similar review of experimental studies into murine subjects showed that, in addition to MRI techniques, invasive methods involving optical measurements of liquid tracers or microspheres motion through cranial windows or thinned skulls have been employed (such as two-phase laser scanning microscopy, two-photon imaging microscopy and Doppler optical coherence tomography)~\cite{Iliff12, Yoshihara13, Bedussi17, Mestre18, Bedussi18, Sweeney19}. While these additional methods cannot be employed on humans because they are invasive, have a limited depth penetration and/or can damage brains through tissue heating, they have provided measurements of various murine perivascular anatomies and associated CSF flows. For example, the experimental consensus is that murine mean axial CSF velocities in PVSs surrounding pial and penetrating arteries are in the region of $10-20 \ \mu$m\,s$^{-1}$~\cite{Mestre18, Bedussi18, Raghunandan21, Kelley21, Boster23flow} and in the same direction as blood flow.

Given that CSF flow is known to be strongly influenced by the cardiac cycle and that blood vessels exhibit elasticity, modelling efforts have focused on describing CSF flows driven by peristaltic forces from arterial wall motion.  These models have included computational fluid dynamics (CFD) approaches~\cite{Bilston03, Daversin-Catty20, Kedarasetti20, Vinje21, Gan24}, analytical approaches based on reduced-order models~\cite{Wang11, Romano20, Coenen21, Gjerde23, Boster24} and porous models utilising Darcy and Darcy-Brinkman equations~\cite{Wang11, Asgari16, Rey18}. Initially it was common to adopt an infinitely long blood vessel surrounded by an infinitely long PVS, however a few recent authors have explored finite lengths and branching (e.g.~\cite{Coenen21, Gjerde23}). To the best of our knowledge, in all prior studies employing lubrication theory the arterial wall is assumed to move as a wave without employing a blood pressure wave as a driving force. The deformation of the outer PVS boundary due to CSF flow has been included in some studies (e.g.~\cite{Romano20, Trevino24, Gan24}), as has the inclusion of a permeable outer PVS (e.g.~\cite{Schley06, Romano20}). Of recent interest has been the effects of non-axisymmetric PVS cross sections (e.g.~\cite{Tithof19, Coenen21, Vinje21, Boster24}) and the significance that Aquaporin-4 channels (AQP4) play in CSF leakage out of the PVS and into brain (e.g.~\cite{Gan24}). Studies of the peristaltic arterial wall motion that have analysed velocities, flow rates or hydraulic resistances often differ in the parameter values employed, which is most likely due to the paucity of human and murine anatomical and physiological data relating to brain penetrating arteries and their associated PVSs.

This work goes beyond previous studies by describing a more complex pumping mechanism for CSF flow in the PVS. It consists of two drivers of CSF flow and employs, in the most part, anatomical parametric data that is entirely based on human studies of PVS-penetrating artery configurations. The first driver is an arterial wall motion induced by a pulsatile blood flow within the artery, where the arterial wall motion is described by the model developed by Zamir~\cite{Zamir12}. The pulsatile blood flow results in an oscillatory wall motion that is attenuated and involves both radial and axial directions. The second driver is an oscillation of the outer PVS boundary, representative of the transfer of blood pulsatility into brain tissue, with the brain expanding and contracting in size almost instantaneously in rhythm with the cardiac cycle~\cite{Wagshul11}. 
Recent studies have measured intracranial brain tissue movement, which can vary between subjects and across the brain, see~\cite{Ince20, Turner20}. As far as we are aware, this is the first study that includes brain pulsations in the modeling of CSF motion in the PVS.

We use lubrication theory to derive the equations of CSF motion and perform a parametric study based by varying key anatomical parameters, with and without brain pulsations. The majority of the parametric values employed here have been taken from literature that describes clinically normal ranges associated with PVS-penetrating artery configurations. The primary focus is to establish whether net axial CSF flow rates can be significantly influenced by brain pulsations in humans. Our results suggest that positive net axial flows are present in all but two of our pathological scenarios and such negative flows become positive when brain pulsations are incorporated. The parameter that drives the largest radial wall motion is found to be Young's modulus for the artery. The larger the amplitude of the oscillations of the outer PVS boundary the larger the magnification.

Researchers continue to debate whether findings from murine studies can be reliably extrapolated to human anatomy and physiology, given the substantial differences between the two species. These differences include disparities in size, variations in physiological mechanisms, and structural distinctions, especially during aging. Further complicating interpretation, murine experiments are often conducted on anaesthetized animals that are physically restrained. Moreover, there are currently no universally recognized standards or scaling factors to bridge murine and human data. Specifically, in relation to CSF flows in PVS, numerous experimenters and modellers emphasise that measurements of CSF velocities, flow rates, and related hydraulic resistances are highly dependent on the specific anatomy being examined~\cite{Thomas19, Gjerde23, Boster24}, there being a significant difference between humans and murines. For these reasons we do not think it is appropriate to present a direct comparison of our results to those of murine studies in the main body of this paper. However, recognising that some mammalian commonalities exist (such as similarities in tissue types and PVS locations) and given the lack of other human studies presenting quantitative CSF flow data, we provide some comparison information for interested readers in the appendix.

The rest of the paper is structured in the following way. The mathematical model is outlined in section \ref{sec:2}, followed by a presentation of model results along with a discussion in section \ref{sec:3}. Conclusions and recommendations are outlined in section \ref{sec:4} and we consider murine data from other studies in an appendix to this paper.

\section{Mathematical model }
\label{sec:2}

\subsection{Geometry and model parameters}\label{subsec:2.1}
\begin{figure}[!ht]
\centering
\includegraphics[width=0.85\textwidth]{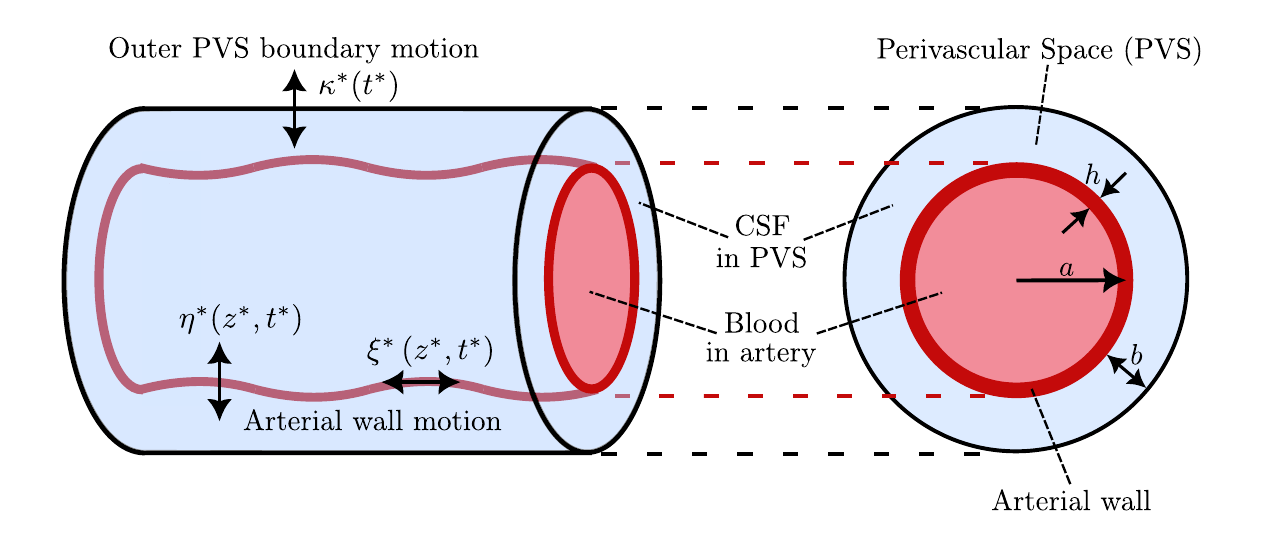}
\caption{The geometry of PVS-penetrating artery anatomy being studied is illustrated. The wall of the penetrating artery (shown in deep red) constitutes the inner boundary of the PVS and moves in two dimensions in accordance with  functions $\eta^* \left( z^*, t^* \right)$ and $\xi^* \left( z^*, t^* \right)$ as a result of a cardiac driven blood pressure wave. The outer PVS boundary moves in accordance with the function $\kappa^* \left( t^* \right)$ as a result of brain pulsations. CSF moves within the PVS and is shown in blue. The neutral radius of the artery is $a$ with a wall thickness of $h$ and the mean PVS thickness is $b$.}
\label{fig: A1}
\end{figure}

The goal of this paper is to model the flow of CSF through the arterial PVS. Here, a simplified geometry is used, as shown in Figure 1. The  PVS  is modeled as the annular gap between an elastic artery and the surrounding brain tissue. Physically the outer PVS boundary is made up of a thin pia membrane behind which brain tissue resides. The brain penetrating artery is characterized by a neutral radius $a$ and wall thickness $h$ (see Fig.~\ref{fig: A1}), and a homogeneous, linearly elastic material defined by density $\rho_w$, Young's modulus $E$ and Poisson's ratio $\sigma$. The PVS is filled with CSF, which is assumed to be an incompressible, Newtonian fluid of viscosity $\mu_c$ and density $\rho_c$, and subject to negligible body forces. 

Driven by the cardiac cycle, blood with density $\rho_b$ and viscosity $\mu_b$ flows through the artery as a longitudinal pressure wave, interacting with the elastic arterial wall and resulting in both radial and axial wall displacements. Here, the approach described in \cite{Zamir12} is adopted to describe such displacements. In particular, by assuming an input oscillatory blood pressure gradient with angular frequency $\omega$ and amplitude $B$, the  displacements in the axial ($z^*$) and radial ($r^*$) directions are given at time $t^*$, respectively, by:
\begin{subequations}
\label{eq: Displacements}
\begin{eqnarray}
        \xi^*(z^*,t^*) &= \mathrm{Re}\left[\hat{\xi}_0\exp\left(i\omega\left(t^*-\dfrac{z^*}{\hat{c}}\right)\right) \right], \label{eq: Displacement xi}\\
        \eta^*(z^*,t^*) & =  \mathrm{Re}\left[\hat{\eta}_0\exp\left(i\omega\left(t^*-\dfrac{z^*}{\hat{c}}\right)\right)\right], \label{eq: Displacement eta} 
\end{eqnarray}
\end{subequations}
where asterisks represent dimensional variables and $\mathrm{Re}[\cdot]$ denotes the real part. The quantities  $\hat{\xi}_0$, $\hat{\eta}_0$, and $\hat{c}$ are complex numbers given  by
\begin{equation}
\hat{\xi}_0 = \frac{i}{\rho_b \omega \hat{c} } \left[\frac{2 - q(1 - g)}{q(2\sigma - g)}\right] B, \qquad
\hat{\eta}_0 = \frac{a}{\rho_b \hat{c}^2} \left[\frac{g + \sigma q(g - 1)}{q(g - 2\sigma)}\right] B,\qquad \hat{c} = \sqrt{ \frac{E h}{\rho_b a ^2q(1-\sigma^2)}},
\end{equation}
which depend on the following quantities:
\begin{equation}\label{displacement-breakdown-appendix}
g = \frac{2 J_1(\Lambda)}{\Lambda J_0(\Lambda)}, \qquad
\Lambda = a\left(\frac{i - 1}{\sqrt{2}}\right)\sqrt{\frac{\rho_b \omega}{\mu_b}},
\end{equation}
where $J_0$ and $J_1$ are the Bessel functions of the first kind of orders zero and one, respectively, and the complex parameter $q$ is a solution of the quadratic equation:
\begin{equation}
    \left[(g-1)(\sigma^2-1)\right]q^2+\left[\frac{\rho_w h}{\rho_b a}(g-1) + \left(2\sigma -\frac12\right)g-2\right]q+\frac{2\rho_w h}{\rho_b a} +g = 0.
\end{equation}
%
%, and $\hat{c}$ are complex numbers that depend on the physical parameters of the system, see Appendix A for more details. 
The wall displacements, $\xi^*(z^*,t^*)$ and $\eta^*(z^*,t^*)$, given in Eqs.~\eqref{eq: Displacements} are characterized by an attenuation length, $z_a$, wave speed $c_s$, and wavelength $\lambda$, which are given as:
\begin{equation}
% z_a = \lvert {\frac{1}{\omega} \mathrm{Im}\left(\frac{1}{\hat{c}}\right)} \rvert, \qquad  c_s = \left[\mathrm{Re}\left(\frac{1}{\hat{c}}\right)\right]^{-1}, \qquad \lambda = \frac{2\pi c_s}{\omega},
z_a = \left\vert \frac{1}{\omega}\left( \mathrm{Im}\left[\frac{1}{\hat{c}}\right] \right)^{-1}\right\vert, \qquad  c_s = \left(\mathrm{Re}\left[\frac{1}{\hat{c}}\right]\right)^{-1}, \qquad \lambda = \frac{2\pi c_s}{\omega},
% z_a = \left\vert \frac{1}{\omega}\left( \mathrm{Im}\left[\hat{c}^{-1}\right] \right)^{-1}\right\vert, \qquad  c_s = \left(\mathrm{Re}\left[\hat{c}^{-1}\right]\right)^{-1}, \qquad \lambda = \frac{2\pi c_s}{\omega},
\end{equation}
where $\mathrm{Im}[\cdot]$ denotes the imaginary part. %and  $\omega$ is the angular frequency of the blood pressure wave. 
Using Eq.~\eqref{eq: Displacement eta}, the radial coordinate of the inner wall of the PVS is defined as:
\begin{equation}
    r_a^*(z^*,t^*) = R_a + \eta^*(z^*,t^*),
\end{equation}
where $R_a = a+h/2$ is the mean radius of the inner wall (see Fig.~\ref{fig: A1}). %, $a$ is the arterial radius and $h$ the arterial thickness (see Fig.~\ref{fig: A1}).
The outer PVS boundary is substantively brain tissue and we propose that its motion is oscillatory in the radial direction, moving towards and away from the artery in synchronisation with brain pulsations. Specifically, we assume that the motion of the outer PVS oscillates in the radial direction as:
\begin{equation}\label{eq: Outer wall}
r_o^*(t^*) = R_o -\kappa^*(t^*),  
\end{equation}
where $R_o = R_a +b$ and $\kappa^*(t^*) =  K\cos(\omega t^*)$, with $K$ being the amplitude of the outer PVS oscillations. Here, $b$ represents the mean PVS thickness. Recent investigations have concluded that brain tissue amplitudes (BTP) can be observed throughout the human brain and vary widely (from $\sim 4$ $\mu$m to $\sim 150$ $\mu$m), see~\cite{Ince20, Turner20} as indicative studies. However, as BTP studies are in their infancy and no studies have focused on PVS areas, it is not yet possible to quantify how much the outer PVS boundary moves. In our formulation, it is assumed the outer PVS boundary moves freely up to 60\%  of the PVS radial thickness, that is $K \leq 0.60 b$. In addition, we assume that local elastic deformations of the brain tissue are small compared to the brain pulsations given by Eq.~\eqref{eq: Outer wall} and occur over a much longer time scale. This can be justified by considering the parameter $\Lambda = G / (\mu_c \omega)$, which is introduced in \cite{Trevino24} and measures the stress balance at the
brain tissue wall. In our case,  we find $\Lambda \sim 10^5$, taking a typical brain tissue shear modulus of $G \sim 2 \times 10^3$ Pa, which is within the rigid limit and so local elastic wall deformations are likely to have only a minor influence under the conditions considered here. 

Table~\ref{tab: parms} outlines the anatomical and physiological parameters employed in the modelling. To ensure clinical relevance, all the values relate to human subjects without pathology and were obtained from imaging journals, histology journals and medical reference books as referenced. The arterial parameters only relate to penetrating arteries and the baseline case is based on studies of the recurrent artery of Heubner, which was chosen as it is one of the largest lenticulostriate arteries and is well classified \cite{falougy2013}. Due to the paucity of studies into PVS-penetrating artery configurations, most of parameter ranges are sourced from single sources.

\begin{table}[!ht]
\centering
\setlength{\tabcolsep}{0pt} 
\begin{tabular}{|l|c|c|l|}
\hline
\cellcolor{gray!25} & \cellcolor{gray!25} & \cellcolor{gray!25} & \cellcolor{gray!25} \\
\cellcolor{gray!25} \emph{Arterial parameters} & \cellcolor{gray!25}\emph{ Baseline value } 
& \cellcolor{gray!25}\emph{Normal Range} & \cellcolor{gray!25}\emph{ References \ {  }} \\
\cellcolor{gray!25} & \cellcolor{gray!25} & \cellcolor{gray!25} & \cellcolor{gray!25} \\
\hline
& & & \\
\ Mean radius, $R_a$ (m) & $3.34 \times 10^{-4}$ & $0.48$--$3.79 \times 10^{-4}$ & ~\cite{Djulejic15} \\
\ Wall-to-lumen ratio, $R_{wl}$ & $7.9 \times 10^{-2}$ & $7.5$--$8.3 \times 10^{-2}$ & ~\cite{Nakagawa16} \\
\ Length, $L$ (m) & $2.4 \times 10^{-2}$ & $2.0$--$5.0 \times 10^{-2}$ & ~\cite{Gomes84, Maga13, Osiowski24} \\
\ Young's modulus, $E$ (Pa) & $4 \times 10^{5}$ & $2$--$6 \times 10^{5}$ & ~\cite{Ebrahimi09} \\
\ Wall density, $\rho_a$ (kg/m$^3$) & $1.102 \times 10^{3}$ & \ $1.056$--$1.147 \ \times 10^{3}$ & ~\cite{itis24} \\
\ Poisson's ratio, $\sigma$ & $0.49098$ & $0.49087$--$0.49109$ & ~\cite{Karimi16} \\
& & & \\
\hline
\cellcolor{gray!25} & \cellcolor{gray!25} & \cellcolor{gray!25} & \cellcolor{gray!25} \\
\cellcolor{gray!25} \emph{Cardiac and blood parameters} 
& \cellcolor{gray!25}\emph{Baseline value} 
& \cellcolor{gray!25}\emph{Normal Range}
& \cellcolor{gray!25}\emph{ References} \\
\cellcolor{gray!25} & \cellcolor{gray!25} & \cellcolor{gray!25} & \cellcolor{gray!25} \\
\hline
& & & \\
\ Heart rate (min$^{-1}$) & $66$ & $60$--$100$ & ~\cite{Dantas18, anatphsiol22, BHF24} \\
\ Blood pressure amplitude, B (mmHg) & $13.5$ & $8.5$--$20$ & ~\cite{Blanco17} \\
\ Viscosity, $\mu_b$ (Pa s) & $3.65 \times 10^{-3}$ & $2.10$--$4.58 \times 10^{-3}$ & ~\cite{itis24} \\
\ Density, $\rho_b$ (kg/m$^3$) & $1.050 \times 10^{3}$ & $1.025$--$1.060 \times 10^{3}$ & ~\cite{itis24} \\
& & & \\
\hline
\cellcolor{gray!25} & \cellcolor{gray!25} & \cellcolor{gray!25} & \cellcolor{gray!25} \\
\cellcolor{gray!25} \emph{CSF parameters} 
& \cellcolor{gray!25}\emph{Baseline value} 
& \cellcolor{gray!25}\emph{Normal Range}
& \cellcolor{gray!25}\emph{ References} \\
\cellcolor{gray!25} & \cellcolor{gray!25} & \cellcolor{gray!25} & \cellcolor{gray!25} \\
\hline
& & & \\
\ Density, $\rho_c$ (kg/m$^3$) & $1.007 \times 10^{3}$ & -- & ~\cite{itis24} \\
\ Viscosity, $\mu_c$ (Pa s) & $0.85 \times 10^{-3}$ & $0.7$--$1.0 \times 10^{-3}$ & ~\cite{Bloomfield98} \\
& & & \\
\hline
\cellcolor{gray!25} & \cellcolor{gray!25} & \cellcolor{gray!25} & \cellcolor{gray!25} \\
\cellcolor{gray!25} \emph{PVS parameters} 
& \cellcolor{gray!25}\emph{Baseline value} 
& \cellcolor{gray!25}\emph{Normal Range}
& \cellcolor{gray!25}\emph{ References} \\
\cellcolor{gray!25} & \cellcolor{gray!25} & \cellcolor{gray!25} & \cellcolor{gray!25} \\
\hline
& & & \\
\ Radial thickness, $b$ (m) & $0.5 \times 10^{-4}$ & $0.5$--$16.7 \times 10^{-4}$ & \ Assumption \\
\ Outer boundary amplitude, $K$ (\% of $b$) \ { } & -- & $0$--$60$\% & \ Assumption \\
& & & \\
\hline       
\end{tabular}
\caption{Model parameters with associated baseline values, ranges, and source references. The rationale for the assumed values is provided in the main text.}
\label{tab: parms}
\end{table}

\subsection{Dimensionless governing equations}
The CSF flow is assumed to be axisymmetric with a velocity field denoted as $\mathbf{u^*} = (u^*,v^*)$, where $u^*$ is the axial component and $v^*$ is the radial component in a cylindrical coordinate system. 
The axisymmetric fluid equations are written using dimensionless variables by choosing the following characteristic scales. The length scales along the radial and axial directions are $b$, the mean PVS thickness, and $\lambda$, the wavelength of the blood pressure wave interacting with the arterial wall, respectively. As the system exhibits oscillatory flow driven by the cardiac cycle with angular frequency $\omega$, we identify the characteristic time scale $1/ \omega$. With these, we define non-dimensional variables (without asterisks) as: 
\begin{subequations}
    \begin{eqnarray}
    r^* =  br, \qquad z^* = \lambda z, && \qquad t^* = \frac{t}{\omega}, \qquad u^* = \lambda\omega u\, \qquad v^* = b\omega v, \\
    p^* =  \frac{\mu \omega\lambda^2}{b^2} p, && \qquad \xi^* = \lambda\xi, \qquad \eta* = b \eta, \qquad    \kappa^* = b \kappa , 
    \end{eqnarray}
\end{subequations}
The pressure scale has been chosen to balance the viscous terms in the governing  equations. Noting that anatomically the PVS thickness is very much smaller than the wavelength of the blood pressure wave, we define the aspect ratio $\epsilon = b/\lambda \ll 1$. Under these conditions, the Reynolds number $Re$ is generally small (in the order of $Re\sim 10^{-6} - 10^{-4}$~\cite{Mestre18}). Hence, we are assuming a regime where fluid inertia is negligible and fluid flow is governed by the Stokes and continuity equations. In the limit of $\epsilon\to 0$, the governing equations reduce to the standard lubrication system:
\begin{subequations}
    \begin{eqnarray}
        \frac{1}{r}\partial_r\left(r\partial_r u\right) & =& \partial_z p, \label{eq:Lub1}\\
\partial_r p & =& 0, \label{eq:Lub2}\\
\frac{1}{r}\partial_r\left(rv\right) +\partial_z u & =& 0,\label{eq:Lub3}
    \end{eqnarray}
\end{subequations}
where the dimensionless asterisk notation is dropped for convenience in the following. The above system of equations is solved imposing the non-slip boundary conditions
\begin{subequations}
\label{eq:BC velocity}   
    \begin{eqnarray}
u &= u_a,\ & v   = v_a\quad \text{at}\ r = r_a,  \label{eq:BC1} \\
u &= 0,\ & v   = v_o\quad \text{at}\   r = r_o,  \label{eq:BC2}    
    \end{eqnarray}
\end{subequations}
and
\begin{equation}
p = p_0\quad\text{at}\  z=0,\quad \text{and}\quad p  = p_L \quad \text{at}\  z=L,    \label{eq:BC5}\\
\end{equation}
where $v_a = \dot{\eta} (z,t)$ and $v_o = \dot{\kappa} (t)$, and $u_a = \dot{\xi}(z,t)$
is the tangential velocity of the arterial wall. We also consider any linear pressure gradient that exists along the PVS, %, say due to natural CSF production, 
$\Delta p = p_L - p_0$. 

\subsection{Lubrication solution}
Integrating Eq.~\eqref{eq:Lub1} and imposing conditions \eqref{eq:BC1} and \eqref{eq:BC2}, we find the axial component of the velocity profile: 
\begin{equation}
u(r,z,t) = -(\partial_z p)\, U + \frac{\ln(r/r_o)}{\ln(r_a/r_o)}\, u_a,
\end{equation}
where the dependence on $z$ and time enters through the boundary movements, $r_a(z,t)$, $r_o(t)$, and $u_a(z,t)$. The function $U(r;r_a,r_o)$ is given by
\begin{equation}
U(r;r_a,r_o) = \frac{r_o^2-r^2}{4}+\frac{\ln(r/r_o)}{\ln(r_a/r_o)}\frac{r_a^2-r_o^2}{4},
\end{equation}
and hence the CSF volume flow rate $Q(z,t)$ is
\begin{equation} 
\label{eq: flow rate}
\frac{Q}{2\pi} =\int_{r_a}^{r_o} u(z,r,t)\, r \, d r =-\alpha(z,t)\,\partial_zp -\beta(z,t) \,u_a,
\end{equation}
where we have defined the coefficients
\begin{equation}
\label{eq: alpha coefficients}
    \alpha(z,t)  = \frac{r_o^4}{16}\left[1-\delta^4+\frac{\left(1-\delta^2\right)^2}{\ln \delta}\right],\quad \qquad 
    \beta(z,t)  = \frac{r_o^2}{4} \frac{1-\delta^2(1-2 \ln \delta)}{ \ln \delta},    
\end{equation}
and the ratio $\delta(z,t) = r_a(z,t)/r_o(t)$.   Equation \eqref{eq: flow rate} shows that the CSF flow inside the PVS is driven by both pressure gradients {(due to the magnification of radial motions of the arterial wall and by the outer PVS boundary motions) and the axial motion along the arterial wall. 

To derive the CSF pressure field, $p(z,t)$, we integrate the continuity equation \eqref{eq:Lub3} along the radial distance and impose the boundary conditions \eqref{eq:BC velocity}. This gives the following partial differential equation:
\begin{equation}
    \label{eq: CSF pressure PDE}
    \partial_z\left[\alpha(z,t)\,\partial_z p\right]+ \gamma(z,t)  = 0,
\end{equation}
where we have defined the coefficient
\begin{equation}
    \label{eq:gamma}
    \gamma(z,t) = r_av_a -r_ov_o + \beta\left(\partial_z u_a - u_a\,\frac{\partial_z\delta}{\delta\ln\delta} \right).  
%    \frac{r_o^2}{4} \frac{1-\delta^2(1-2\ln\delta)}{\ln^2\delta}\left(\ln\delta\, \partial_z u_a - u_a\,\frac{\partial_z\delta}{\delta}  \right).
    %
    %r_av_a-r_0v_0 - \beta\left(\partial_z u_a - \frac{u_a}{\delta\ln\delta}\partial_z\delta\right)
\end{equation}
Equation \eqref{eq: CSF pressure PDE} cannot be solved analytically. The following subsection describes the numerical approach to compute the pressure field as well as the mean CSF flow rate, which will be analysed in Section \ref{sec:3}.

\subsection{Mean axial flow rate - Numerical algorithm}

To find the pressure profile the equation is recast as a set of two first-order ordinary differential equations and solved numerically as a two point boundary value problem (BVP). In all our work, we take $p_0=p_L=0$ in Eqs.~\eqref{eq:BC5} to restrict analysis to the oscillatory motion of the CSF. The axial domain of the PVS, $[0,L/\lambda]$, is discretised into 1000 mesh points and we use a standard Python SciPy library function to solve Eq.~\eqref{eq: CSF pressure PDE} at a given time. The function employs a 4th order collocation algorithm with a standard control of residuals and a damped Newton method with an affine-invariant criterion function. Results were carefully checked for finite size effects. For the baseline parameter values outlined in Table~\ref{tab: parms}, the wavelength of the arterial wall oscillations is $\lambda \approx 0.2$ m and so the dimensionless domain size is $L/\lambda = 0.12$.

Once the pressure profile, $p(z,t)$, is found numerically at a given time, we use Eq.~\eqref{eq: flow rate} to find the volume flow rate, $Q(z,t)$. This process is done over a cardiac cycle to find the time-dependence of both the pressure and flow rate. In particular, we discretize time into 480 time points spanning a cardiac cycle. From the instantaneous flow rate we calculate the flow rate averaged over the length of the PVS:
\begin{equation}
    \overline{Q}(t) = \frac{\lambda}{L}\int_0^{L/\lambda} Q(z,t)\,dz,
    \label{eq: average z flow rate}
\end{equation}
and the mean flow rate  over the length of the PVS and over a cardiac cycle as:
\begin{equation}
    \langle Q\rangle = \int_0^1 \overline{Q}(t)\,dt.
    \label{eq: mean flow rate}
\end{equation}
The dependence of $\langle Q\rangle$ on the different parameters of the model will be analyzed in Section \ref{sec:3} below. 

\begin{figure}[t]
\centering
\includegraphics[width=0.98\textwidth]{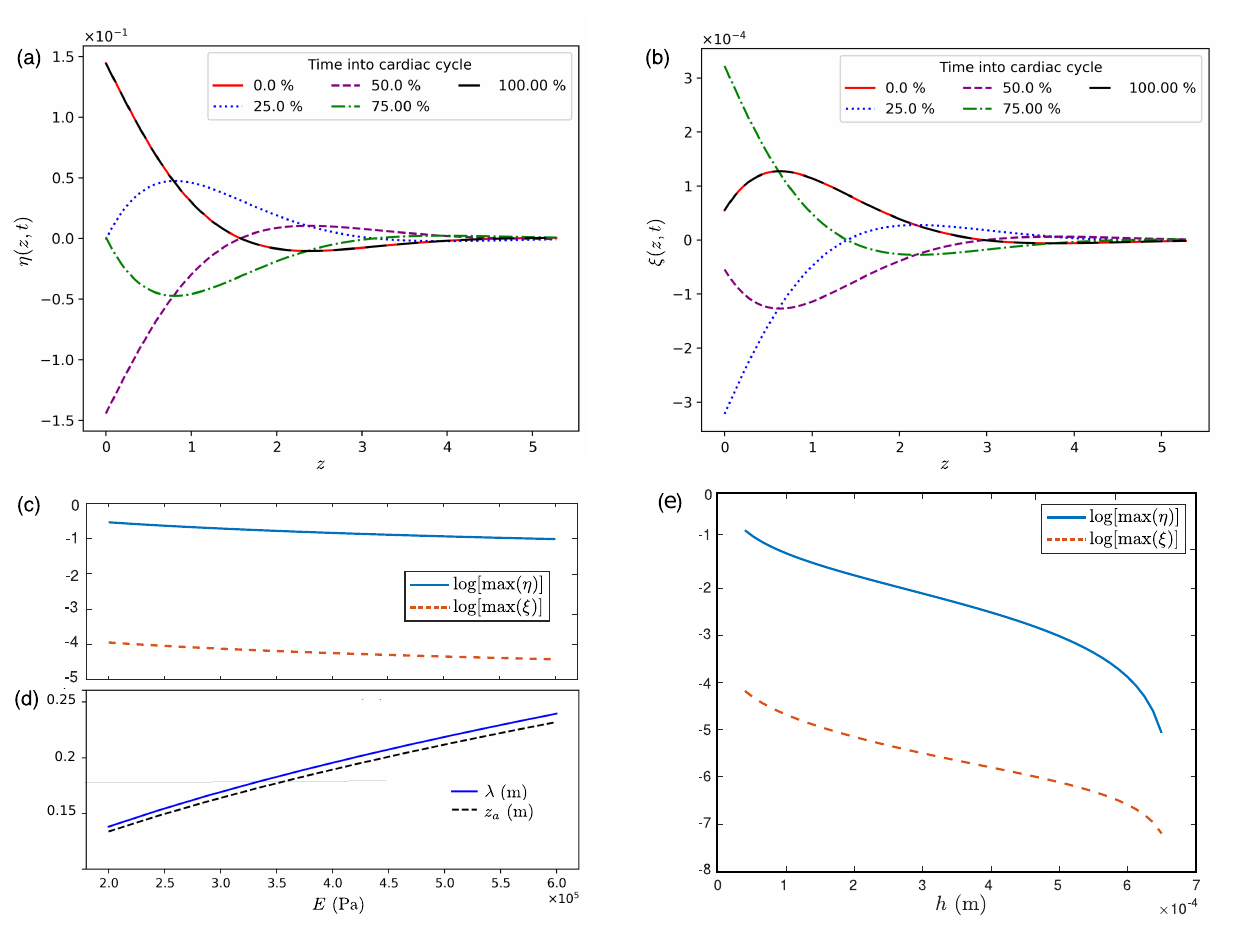}
 \caption{Panels (a) and (b) show the dimensionless radial displacement, $\eta(z,t)$, and the dimensionless axial displacement, $\xi(z,t)$, respectively, plotted against the dimensionless axial length ($z$) at five time points spanning the cardiac cycle. These results correspond to the baseline parameter values outlined in Table~\ref{tab: parms}. In all cases, displacements are shown over a distance significantly greater than the length of the artery used in the baseline model, $L/\lambda \approx 0.12$. 
 Panel (c) shows the maximum dimensionless displacements along the axial length and over a cardiac cycle as a function of the Young's modulus. Panel (d) presents the variation of the wavelength, $\lambda$ (blue solid line), and attenuation length, $z_a$ (black dashed line), with respect to the Young's modulus. Panel (e) displays the maximum dimensionless displacements of $\eta(z,t)$ and $\xi(z,t)$ along the axial length and over a cardiac cycle as a function of a large pathological range of the arterial wall thickness, $h$.
}
\label{fig: A2}
 \end{figure}

\subsection{Arterial wall motion}
Before examining CSF flow rates, it is important to highlight the key properties of the arterial wall model used in this work. Here, we adopt the arterial wall displacements developed by Zamir~\cite{{Zamir12}} to model arterial wall motion. This differs from all previous studies on this problem employing lubrication theory, where the arterial wall is assumed to move as a wave without employing a blood pressure wave as a driving force (see, e.g.~\cite{Romano20, Coenen21}). Equations \eqref{eq: Displacements} provide a more detailed representation of arterial wall motion by incorporating the interaction between a sinusoidal longitudinal blood pressure wave and an elastic arterial wall. In our study, we have assumed that arterial wall deformations are mainly driven by pressure oscillations within the artery, and there is no coupling between the artery wall deformations and the motion of the CSF in the PVS. This is a reasonable approximation within the lubrication framework used to derive the CSF flows, considering that the viscosity of blood is much higher than that of the surrounding CSF. Under these conditions, the coupling between the arterial wall and the CSF along with brain tissue oscillations is expected to have a minimal effect on arterial wall deformation.

The displacements of the arterial wall over a cardiac cycle are illustrated in Figure \ref{fig: A2}. Panels \ref{fig: A2}(a) and \ref{fig: A2}(b) display the dimensionless arterial wall displacements, $\eta(z,t)$ and $\xi(z,t)$, respectively, for the baseline parameter values outlined in Table~\ref{tab: parms}. Both displacements exhibit traveling oscillations that attenuate along the axial distance. Axial wall displacements appear three orders of magnitude smaller in nondimensionalised terms when compared to the radial displacements. While they are one order of magnitude larger than the radial displacements in physical terms, it is important to recognise that the dimensional scaling factors are significantly different, with radial motion scaled by $b=0.5 \times 10^{-4}$ m and axial motion scaled by $\lambda \approx 0.2$ m. We also note that the phase difference between the radial and tangential deformations arises from the fact that the amplitudes $\hat{\xi}_0$ and $\hat{\eta}_0$ in Eqs.~\eqref{eq: Displacements} are complex quantities with different dependencies on the model parameters.

For this parameter set, the wavelength of the traveling wave is $\lambda \approx 0.2$ m, significantly larger than the PVS thickness ($b = 50 \ \mu$m). The attenuation length is of similar order, $z_a \approx 0.19$ m, and much greater than the arterial length ($L = 2.4 \times 10^{-2}$ m). Consequently, in the results presented in the next section, the pumping effect experienced by the CSF in a penetrating artery does not attenuate much over the length of the domain.

By varying individual parameter values within the clinically normal ranges identified in Table~\ref{tab: parms} and studying the corresponding motion of the arterial wall, we find that Young's modulus  has the greatest influence on wall displacements. Panel \ref{fig: A2}(c) shows the maximum arterial wall deformations along the axial length and over a cardiac cycle as a function of the Young's modulus. We observe that both radial and axial wall displacements decrease similarly as the Young's modulus increases, indicating that lower Young’s modulus values result in greater arterial displacements. In addition, panel \ref{fig: A2}(d) shows both the attenuation length, $z_a$, and wavelength, $\lambda$, increase with the Young's modulus.

Panel \ref{fig: A2}(e) presents the maximum arterial wall deformations along the axial length and over a cardiac cycle as a function of the arterial wall thickness, $h$. For this plot, we use a range of $h$ values for the recurrent artery of Heubner (our baseline case) that extends beyond the physiological range to include pathological arterial wall thicknesses (e.g., in cases of arteriosclerosis). We find that within the normal range, the maximum displacements decay similarly, but for $h > 4\times 10^{-4}$ m, the amplitude of the radial displacement decreases much more rapidly than that of the axial displacement.

\section{Results: CSF flow rates}
\label{sec:3}

In this section the model's predictions for mean axial CSF flow rates are presented for the clinically normal ranges of the anatomical parameters listed in Table~\ref{tab: parms}. In addition, two scenarios are considered with extended parameter ranges associated with pathological conditions.

Figure~\ref{fig: B1} shows the dimensionless flow rate calculated using the baseline parameter values outlined in Table~\ref{tab: parms}. Panel~(a) presents the instantaneous flow rate $Q(z,t)$, given by Eq.~\eqref{eq: flow rate}, plotted against $z$ at five points spanning the cardiac cycle. The results show that the flow rate oscillates between a linear and a non-linear dependence on $z$ over a cardiac cycle.
Panel~(b) shows the averaged flow rate $\overline{Q}(t)$, given by Eq.~\eqref{eq: average z flow rate}, for different values of the parameter $K/b$, which controls the amplitude of the brain tissue oscillations, $\kappa(t)$, as defined in Eq.~\eqref{eq: Outer wall}. The time oscillations of $\overline{Q}(t)$ are amplified as $K/b$ increases.

\begin{figure}[!t]
\centering
\includegraphics[width=0.95\textwidth]{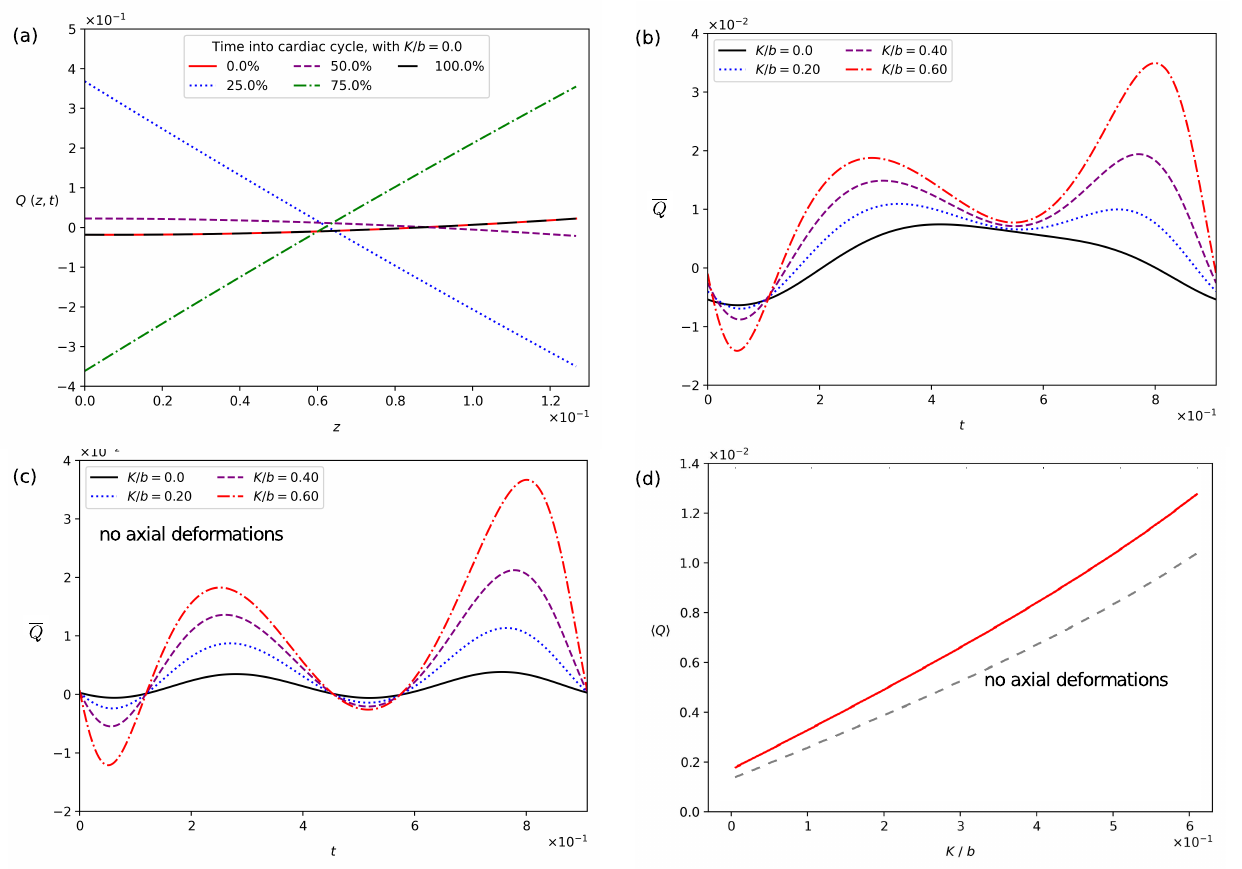}
\caption{Dimensionless flow rates calculated using the baseline parameter values outlined in Table I. Panel (a) shows the instantaneous flow rate $Q(z,t)$ plotted against the dimensionless length $(z)$ at five points spanning the cardiac cycle and with no brain oscillations, $K/b=0$.  Panels (b) and (c) show the averaged flow rate $\overline{Q}(t)$, Eq.~\eqref{eq: average z flow rate}, with and without arterial axial deformations, respectively. In both cases, different brain oscillation amplitudes are considered. Panel (d) shows the mean flow rate $\langle Q\rangle$, Eq.~\eqref{eq: mean flow rate}, as a function of the amplitude of the brain tissue oscillations, $K/b$. The dashed line corresponds to the case with no arterial axial deformations.}
    \label{fig: B1}
\end{figure}

To illustrate the effect of arterial axial deformations, quantified by $\xi(z,t)$, panel~(c) shows the same averaged flow rate as in panel~(b) but with $\xi = 0$. In this case, the removal of axial deformations produces a more sinusoidal oscillation, with the averaged flow rate remaining closer to zero.
In panel~(d), we analyse the effect of varying the amplitude of brain tissue oscillations on the dimensionless mean axial CSF flow rate, $\langle Q\rangle$, as defined in Eq.~\eqref{eq: mean flow rate}. Since our model is not designed to account for occlusions (i.e. PVS collapse), we restrict $K/b$ to values up to 0.6 to ensure such scenarios do not occur during the numerical analysis.

Our results show that brain pulsations significantly enhance the arterial pumping mechanism, increasing the mean axial flow rate. In dimensional terms, these correspond to mean axial CSF flow rates of $6$–$42\times 10^6\ \mu\mathrm{m}^3\,\mathrm{s}^{-1}$, with associated mean axial CSF velocities of $40$–$390 \ \mu\mathrm{m}\,\mathrm{s}^{-1}$. As no experimental measurements of such velocities or flow rates in humans could be found, the only available comparisons are with models that use human parameters in their simulations. Few such models exist, and they predict values in the range $1$–$30 \ \mu\mathrm{m}\,\mathrm{s}^{-1}$~\cite{Daversin-Catty20, Gjerde23, Trevino24}. Our findings are broadly consistent with these predictions, particularly as brain pulsations extend the previously estimated range.
Figure~\ref{fig: B1}(d) also highlights the effect of the arterial axial deformations: removing them results in an overall decrease in flow rate (dashed line).

The relationships between mean flow rates and cardiac heart rate, pulsatile blood pressure amplitude, and arterial radius, are found to be positive monotonic as shown in panels (a), (b), and (c) of Fig.~\ref{fig: B2} respectively. In all three cases, larger parameter values lead to higher flow rates, significantly enhanced by brain pulsations. Panel \ref{fig: B2}(c), in particular, illustrates that larger arteries, with correspondingly greater radii, generate higher flow rates than smaller ones. Brain pulsations further amplify this effect, following a trend similar to that observed for heart rate and blood pressure variations (panels \ref{fig: B2}(a) and \ref{fig: B2}(b)).

\begin{figure}
\centering
\includegraphics[width=0.98\textwidth]{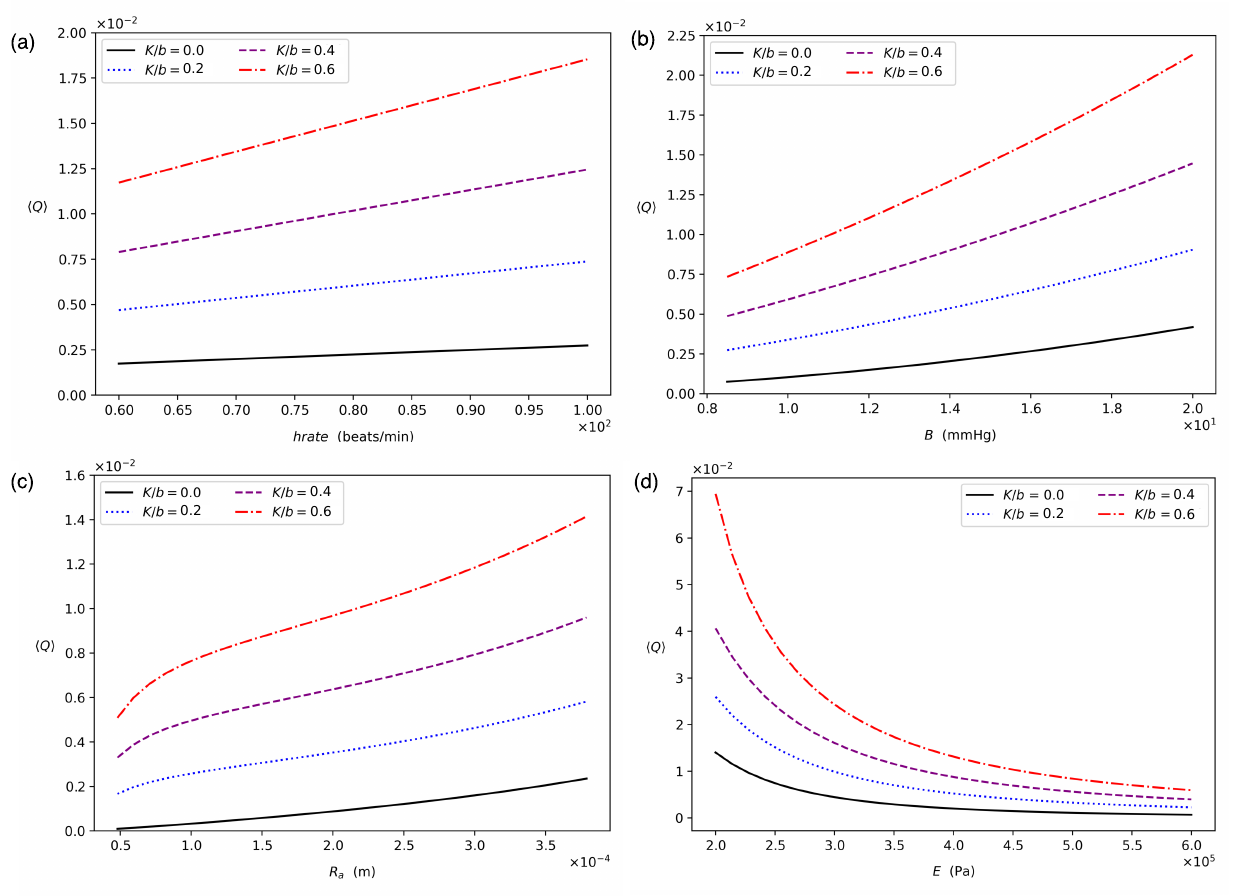}
\caption{Dimensionless mean axial CSF flow rate, $\langle Q\rangle$, as a function of (a) heart rate, $\left( hrate \right)$, (b) blood pressure wave amplitude, $\left( B \right)$, (c) arterial radii, $\left( R_a \right)$, and (d) arterial wall’s Young’s modulus ($E$).  In all cases, different brain oscillation amplitudes are considered: $K/b = 0$ (solid black line), $0.2$ (dotted blue line), $0.4$ (dashed purple line), and $0.6$ (dashdotted red line). 
}
    \label{fig: B2}
\end{figure}

\begin{figure}
\centering
\includegraphics[width=0.98\textwidth]{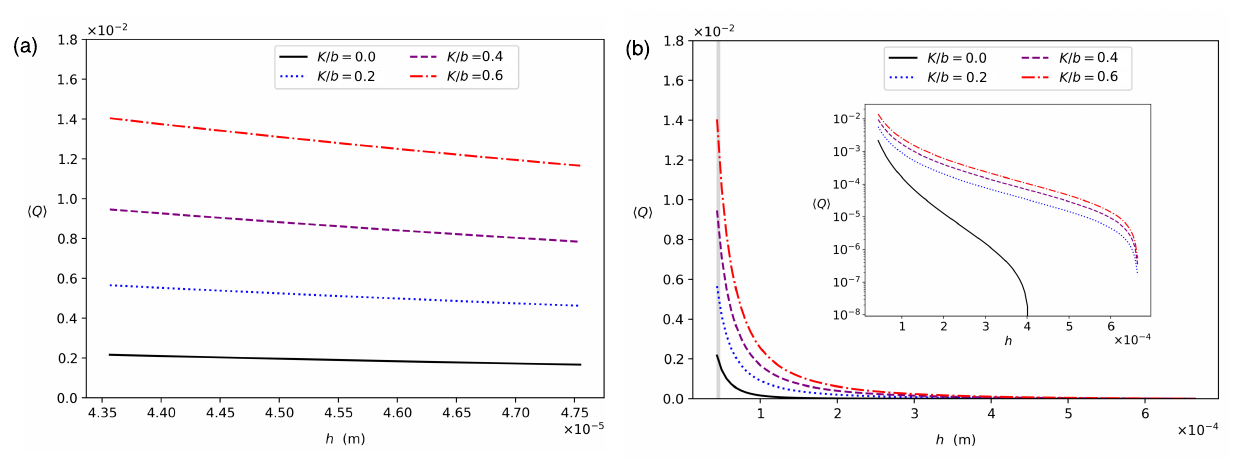}
    \caption{(a) Dimensionless mean axial CSF flow rates, $\langle Q\rangle$, as a function of arterial wall thickness $\left( h \right)$. Panel (a) shows results for the clinically normal range. Panel (b) shows that the mean axial CSF flow rate falls off steeply as the arterial wall thickness $\left( h \right)$ moves outside the clinically normal range, which is depicted by the grey band. In all cases, different brain oscillation amplitudes are considered: $K/b = 0$ (solid black line), $0.2$ (dotted blue line), $0.4$ (dashed purple line), and $0.6$ (dashdotted red line). When no brain pulsations are included relatively very small negative flow rates are observed, as shown in the semi-log plot of the inset of panel (b).}
    \label{fig: B3}
\end{figure}

The effect of the arterial wall’s Young’s modulus on the mean flow rate is negative monotonic, as illustrated in panel (d) of Fig.~\ref{fig: B2}. To our knowledge, no studies have reported a Young’s modulus value for the recurrent artery of Heubner (our baseline case) or established a direct link between specific PVS-penetrating artery configurations and the Young’s modulus value. Therefore, we assume a baseline value corresponding to the midpoint of the general range of Young's moduli for arteries reported in \cite{Ebrahimi09} and focus on examining the impact of varying brain pulsation strengths.

During the natural aging process arterial walls not only thicken, but also a process of arterial remodelling takes place, which involves a reduction in elastin and an increase in collagen in the arterial wall composition~\cite{Fonck09, Castelli23}. The effect of this remodelling is that arteries become more rigid. The results in Fig.~\ref{fig: B2}(d) indicate  that higher Young’s modulus values (greater rigidity) correspond to lower flow rates, confirming the expectation  that a less elastic artery generates weaker peristaltically driven CSF flow.  When brain pulsations are included in the model, the CSF flow rates increase, however the rate at which they decrease  with increasing rigidity remains unchanged.

\begin{figure}
\centering
\includegraphics[width=0.98\textwidth]{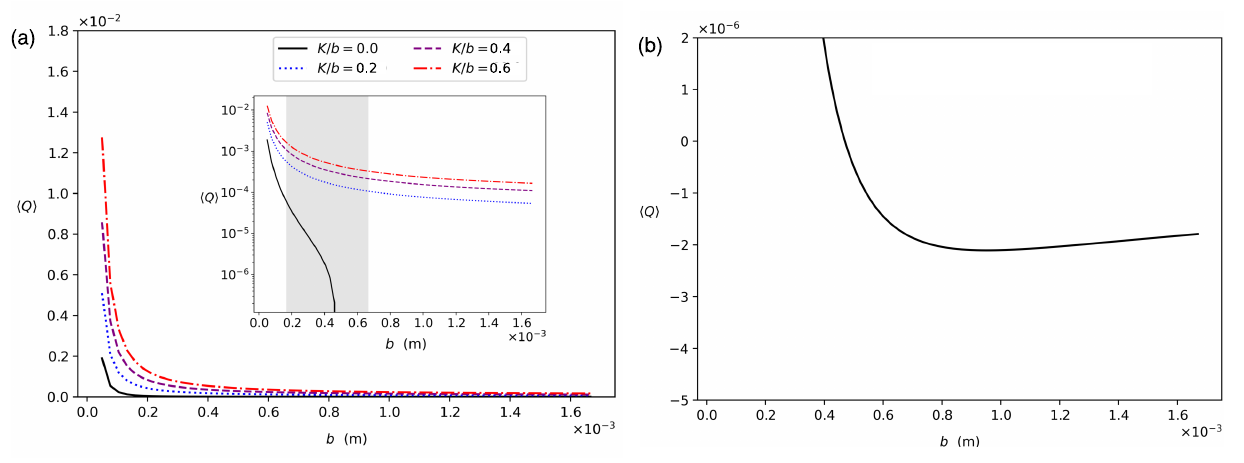}
    \caption{(a) Dimensionless mean axial CSF flow rate, $\langle Q\rangle$, as a function of the PVS thickness $\left( b \right)$. The lowest value of $b$ is the baseline case, and the highest value of $b$ equates to a PVS-penetrating artery configuration that appears as 4 mm in diameter on an MRI scan. The grey shading is the typical clinical detection range, where diameters of $1 - 2 \ mm$ are deemed to be the start of pathological interest. Different brain oscillation amplitudes are considered: $K/b = 0$ (solid black line), $0.2$ (dotted blue line), $0.4$ (dashed purple line), and $0.6$ (dashdotted red line). (b) Negative mean CSF flow rates $\langle Q \rangle$ are possible only when PVS-penetrating artery configuration appears approximately 1.5 mm in diameter on an MRI scan and no brain pulsations exist $K/b = 0$.}
    \label{fig: B4}
\end{figure}

Figure \ref{fig: B3} presents the simulation results for differing values of the thickness of arterial walls within a clinically normal range (panel (a)) and a pathological range (panel (b)). The former is derived from the wall-to-lumen range in Table~\ref{tab: parms}. The latter extends from this range to approximately twice as large as the baseline arterial radius, which is not uncommon in clinical observations. We observe that the flow rate in the clinically normal range decreases only marginally with increased wall-to-lumen ratios. However, increases in arterial thickness in the pathological range are marked by significant decreases in flow rates, even when magnified by brain pulsations. While flows without brain pulsations can be  negative in the pathological range, such flows are very small when compared to the normal positive flows, being approximately five orders of magnitude less and constituting negligible CSF movement. 

Of equal clinical interest, our results confirm the intuitive expectation that mean axial CSF flow rates are lower when PVS thickness is greater, see Fig.~\ref{fig: B4}(a), which shows the mean flow rate as a function of the PVS thickness, $b$. We can see that the mean flow rate falls off steeply as the PVS thickness increases. The lowest value of $b$ considered in this panel is the baseline case, and the highest value of $b$ equates to a PVS-penetrating artery configuration that appears as 4 mm in diameter on an MRI scan. The grey shading shown in panel (a) is the typical clinical detection range, where diameters of $1-2\,$~mm are deemed to be of pathological interest. 

A surprising result from our simulations is the prediction of small negative flow rates under certain pathological conditions. Specifically, negative mean CSF flow rates can occur when no brain oscillations are present, 
as shown in Figs.~\ref{fig: B3}(b) and \ref{fig: B4}(b). We argue that this effect can be induced by a relative decrease in the arterial radial deformations compared to the axial deformations. Figure~\ref{fig: A2}(e) shows that the maximum amplitude of the radial deformation decreases much faster than that of the axial deformations as $h$ increases into a pathological range. Similarly, increasing pathological $b$ values also led to a relative reduction in radial deformations. 
%\sout{Hence, this suggests that the negative flow rates observed at either large $h$ or large $b$ are driven by axial deformations, which become more dominant in these regimes. This interpretation is consistent with the results reported in \cite{Trevino24}. However, it is important to emphasize that }} 
While flow rates can decrease by more than two orders of magnitude in some pathological situations with brain pulsations present, they do not become negative; as in the case of a pathological PVS thickness that means MRI records a recurrent artery of Heubner PVS diameter of over 4 mm. Given that mean flow rates remain consistently positive when brain oscillations are incorporated, our results suggest that backflow may be induced by a relative decrease in radial deformations only when no brain pulsations are present.

As a summary plot of the results presented in this section, Fig.~\ref{fig: B5} provides a comparative view of the impact of each parameter on the mean flow rate. The data is presented with the observation that the relationship between $\langle Q \rangle$ and each parameter is generally monotonic, as seen in all but one case.  The exception lies outside a clinically normal range and relates to the pathological PVS thickness considered in Fig.~\ref{fig: B4}.

\begin{figure}
    \includegraphics[width=0.7\textwidth]{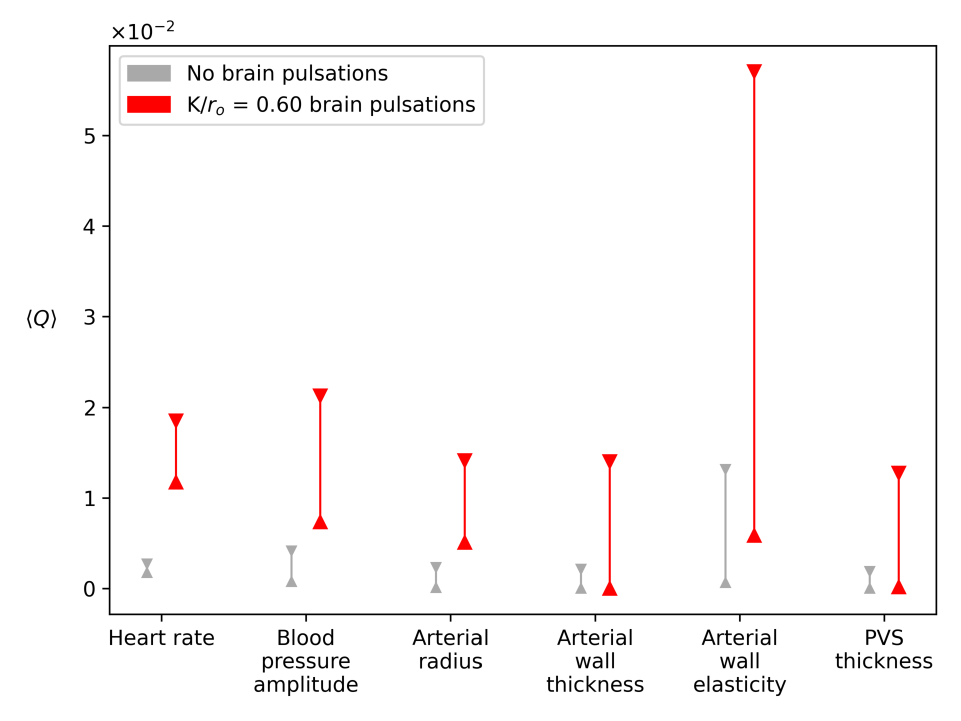}
    \caption{The relative effects of varying parameters on dimensionless mean axial CSF flow rates, $\langle Q\rangle$, with no brain pulsations (in grey) and brain pulsations with amplitude $K/b=0.6$ (in red). Downward pointing triangles results for maximal end of the range and upward pointing triangles the minimum end of the range, with the lines between them a guide to the eye.}
    \label{fig: B5}
\end{figure}

\section{Discussion and Conclusions}
\label{sec:4}

This paper investigated the combined effects of oscillating outer PVS boundaries, representing brain pulsations, and arterial wall motion, from a blood pressure wave, on CSF flow. Lubrication theory was used to create a mathematical model of these effects. Parametric analysis was undertaken to investigate the effect of the brain pulsations and other clinically relevant parameters. To our knowledge, this is the first time that brain pulsations have been modeled as a driver of CSF flow.

Overall, our results show that no one single factor causes increase or decrease in CSF flow. The model predicts that some of the PVS-penetrating artery  parameters are more significant than others when it comes to the motion of the arterial wall. Results also suggest that CSF flow rates are significantly magnified by brain pulsations. While the  hypothesized glymphatic system is not yet been fully elucidated, we conclude that brain pulsations could be a significant missing piece in the complicated brain waste clearance puzzle.  

We briefly discuss the interplay between clinical scenarios and changes in key model parameters, that could modify mean CSF flow and potentially waste clearance. 
\begin{itemize}
\item \textbf{Hypertension, arterial remodeling, arteriosclerosis and vessel elasticity:} While our model indicates that higher blood pressure produces larger CSF flow rates, we note that prolonged hypertension is associated with arterial remodeling~\cite{Fonck09, Castelli23} with resulting lowering of arterial elasticity. Conditions such as arteriosclerosis also result in greater arterial rigidity~\cite{Mitchell20}. Given our model indicates that reduced arterial elasticity has a stronger effect on CSF flows than higher blood pressure, this could be a stronger contributing factor to neurodegenerative decline. Hypertension is a known, avoidable, major risk factor for dementia in older age \cite{Livingston_Lancet2024}, so improving modelling of its potential consequences for glymphatic clearance of abnormal proteins and macromolecules is important to digital avatar and digital twin studies of dementia being developed to facilitate improved personalized care.

\item \textbf{Brain atrophy and PVS space:} Numerous studies have linked Alzheimer's and Parkinson's diseases to the greater prevalence of enlarged PVS visible in MRI scans (e.g.~\cite{Brown22, Zdanovskis22, Yang23}). Similarly, cognitive decline in later life (e.g.~\cite{Schaie90, Wahl19, Paradise21}) is also characterised by increases in PVS thickness (e.g.~\cite{Lynch23, Kim23, Park23}). As brain volume shrinks, the PVS space increases and our model predicts CSF flow reductions, which may account for such links. Furthermore, as initial stages of neurodegenerative conditions can lead to decreases in brain volume, our model would suggest that the resultant reduction in CSF flow (and associated clearance rates) could add to a further neurodegenerative decline. Given the rapid fall-off rate of CSF flows with increased PVS space, we speculate that this could lead to an apparent sudden rapid decline once a critical threshold is reached. 
\item \textbf{Exercise and heart rate:} We have shown that an increase in heart rate can increase flow of the CSF within the PVS and therefore clearance of waste products through the glymphatic system could be increased. This suggests a possible role for exercise in CSF clearance (e.g.~\cite{Andel08, He17}), with the caveat that glymphatic clearance has been thought by some researchers to be aided by sleep where no exercise is undertaken~\cite{Chong22, Sangalli23}. 
\end{itemize}

The results of this work suggest several future directions. Measurements in outer PVS boundary oscillations in humans \textit{in-vivo} would be useful to confirm the relevance of this model and its clinical significance, once technological advances permit.  Further modeling could be undertaken to explore the effects of brain pulsations on CSF in the subarachnoid space (SAS) and how this contributes to CSF flows in the PVS. In addition, it would be insightful to quantify any link between brain tissue oscillations and blood pressure and how this could be incorporated into a mathematical model. Improvements to incorporate recent scientific understanding of arterial wall aging into the model would also be beneficial, as would introducing a more realistic mammalian blood pressure wave equation.

Finally, further study of the interaction between CSF flow rates and AQP4 channels would be warranted. While the inclusion of a permeable outer PVS has been considered for some time (e.g.~\cite{Schley06, Romano20}), interest is growing in the significance that AQP4 channels play in the glymphatic system (e.g.~\cite{Gan24}). Our model could be extended to examine radial CSF flows in PVS and their relationship to AQP4 channel expressions.

\section*{Acknowledgments}

We are grateful to the anonymous referees for valuable comments and critical suggestions. We thank the Engineering and Physical Sciences Research Council, UKRI for funding our work under grant EP/W52458X/1 and for the support of The Open University and the University of Sheffield in the UK.

\appendix*

\section{Murine data in other studies}
\label{sec:5}
A time-boxed review of anatomical and physiological murine data was conducted with the view to extract comparable results. While it was easier to identify a larger body of murine studies than human ones, it proved harder than anticipated to identify data that could be used as direct comparisons to our work as the studies covered different species (mice and rats), different arterial blood vessels types (e.g. MCA, PCA, Pial, penetrating) and even different PVS locations (e.g. spinal chord, SAS or parenchyma).

The murine studies most comparable to our work investigate CSF flow in perivascular spaces (PVSs) surrounding both penetrating arteries and pial arteries; the latter being larger surface vessels from which penetrating arteries branch. Experimental results from these studies indicate mean axial CSF velocities in the range of approximately $10-20 \ \mu$m\,s$^{-1}$~\cite{Mestre18, Bedussi18, Raghunandan21, Kelley21, Boster23flow}.  In comparison, the baseline mean axial CSF velocities measured in our human study ($40$–$390 \ \mu$m\,s$^{-1}$) are substantially higher than those reported in experimental murine work.

\begin{figure}[!t]
\centering
\includegraphics[width=0.55\textwidth]{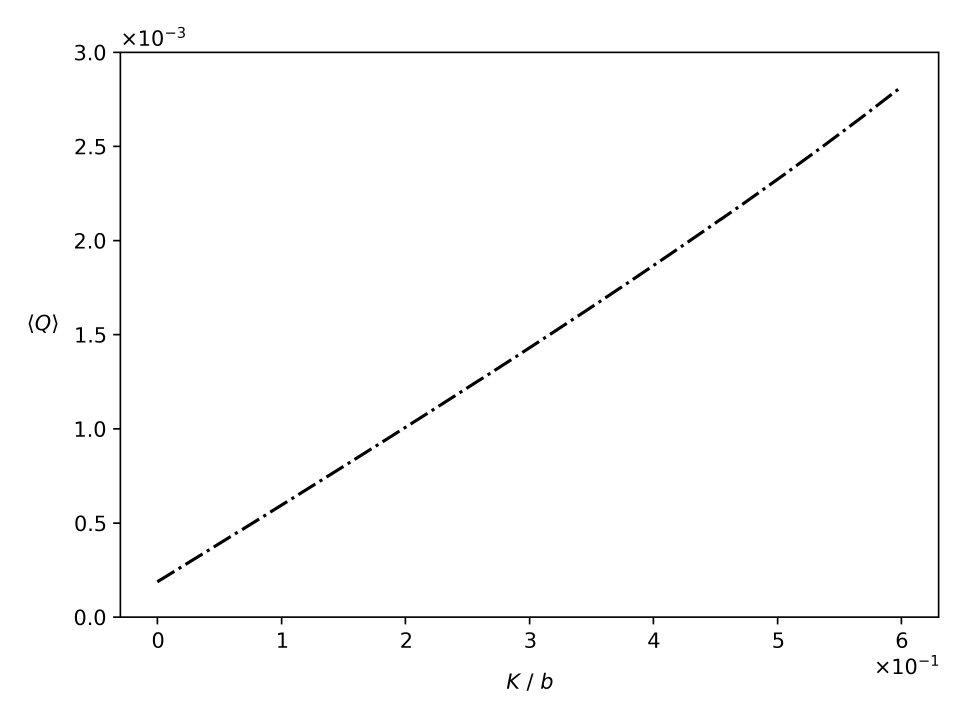}
\caption{Dimensionless mean axial CSF flow rate predicted by the model for mice data, $\langle Q\rangle$, as a function of the amplitude of the brain tissue oscillations, $K/b$.}
    \label{fig: 8-mice}
\end{figure}

\begin{table}[!t]
\centering
\setlength{\tabcolsep}{0pt} 
\begin{tabular}{|l|c|l|}
\hline
\cellcolor{gray!25} & \cellcolor{gray!25} 
& \cellcolor{gray!25} \\
\cellcolor{gray!25} \emph{Arterial parameters} & \cellcolor{gray!25}\emph{ Baseline value } & \cellcolor{gray!25}\emph{ References and rationale \ {  }} \\
\cellcolor{gray!25} & \cellcolor{gray!25} 
& \cellcolor{gray!25} \\
\hline 
& & \\
\ Mean radius, $R_a$ (m) & $ 10.8\times 10^{-6}$ & \ Averaged from~\cite{Iliff13, Sekiguchi13, Yoshihara13,  Unekawa17,  Bojarskaite23}.\\
\ Wall-to-lumen ratio, $R_{wl}$ & $ 1.4 \times 10^{-2}$ &  \ Lowest realistic value from young mice data~\cite{Diaz-Otero16}. \\
\ Length, $L$ (m) & $134  \times 10^{-6}$ & \ Averaged from studies by~\cite{Yoshihara13}. \\
\ Young's modulus, $E$ (Pa) & $4.0 \times 10^{5}$ & \ No studies identified; assumed to be identical to human.\\
\ Wall density, $\rho_a$ (kg/m$^3$) & $1.102 \times 10^{3}$  & \ No studies identified; assumed to be identical to human. \\
\ Poisson's ratio, $\sigma$ & $0.49098$ & \ No studies identified; assumed to be identical to human.\\
& & \\
\cellcolor{gray!25} & \cellcolor{gray!25} 
& \cellcolor{gray!25} \\
\cellcolor{gray!25} \emph{Cardiac and blood parameters} & \cellcolor{gray!25}\emph{ Baseline value } & \cellcolor{gray!25}\emph{ References and rationale \ {  }} \\
\cellcolor{gray!25} & \cellcolor{gray!25} 
& \cellcolor{gray!25} \\
& & \\
\ Heart rate (min$^{-1}$) & $600$ &  \ Midpoint of 500-700 cited by~\cite{Rohrer98,  Kass98, Janssen04, Ho11}. \\
\ Blood pressure amplitude, B (mmHg) & $10.0$ & ~\cite{de-Montgolfier20} cites~\cite{Baumbach02} as quoting 9 mmHG for pial PVSs.\\
\ Viscosity, $\mu_b$ (Pa s) & $3.0 \times 10^{-3}$ & \ High shear range from~\cite{Ruengsakulrach07}.  \\
\ Density, $\rho_b$ (kg/m$^3$) & $1.050 \times 10^{3}$ & \ Assumed to be identical to human. \\
& & \\
\hline
\cellcolor{gray!25} & \cellcolor{gray!25} 
& \cellcolor{gray!25} \\
\cellcolor{gray!25} \emph{CSF parameters} & \cellcolor{gray!25}\emph{ Baseline value } & \cellcolor{gray!25}\emph{ References and rationale \ {  }} \\
\cellcolor{gray!25} & \cellcolor{gray!25} 
& \cellcolor{gray!25} \\
\hline
& & \\
\ Density, $\rho_c$ (kg/m$^3$) & $1.007 \times 10^{3}$ & \ Assumed to be identical to human.\\
\ Viscosity, $\mu_c$ (Pa s) & $0.85 \times 10^{-3}$ & \ Assumed to be identical to human.\\
& & \\
\hline
\cellcolor{gray!25} & \cellcolor{gray!25} 
& \cellcolor{gray!25} \\
\cellcolor{gray!25} \emph{PVS parameters} & \cellcolor{gray!25}\emph{ Baseline value } & \cellcolor{gray!25}\emph{ References and rationale \ {  }} \\
\cellcolor{gray!25} & \cellcolor{gray!25} 
& \cellcolor{gray!25} \\
\hline
& & \\
\ Radial thickness, $b$ (m) & $9.33 \times 10^{-6}$ & \ Averaged from~\cite{Lightfoot73, Rey18, Bojarskaite23}. \\
& & \\
\hline       
\end{tabular}
\caption{Model parameters for mice subjects, along with source references and any rationale employed in deduced values.}
\label{tab: parms-mice}
\end{table}

Previous studies indicate that simulations based on mouse data yield mean CSF velocity values spanning a relatively wide range $10^{-3}-10^{2} \ \mu$m\,s$^{-1}$~\cite{Rey18, Kedarasetti20, Yokoyama21}, most probably reflecting the challenge of establishing consistent anatomical and physiological model parameters. To address this variability, we assembled a set of baseline parameter values specific to mice and ran a simulation using this data in our model. Several parameters were taken directly from the literature, while others were inferred. Table \ref{tab: parms-mice} lists these values, along with their sources and the rationale for any inferred estimates.

Comparing the resultant dimensionless mean axial flow rates (Fig.~\ref{fig: 8-mice}), with our human study (Fig.~\ref{fig: B1}) confirms that our model predicts mice flow rates are lower than human flow rates. An almost linear trend is observable in both relationships, however the mouse simulation exhibits a significantly larger gradient. In physical terms, our predicted dimensional flow rate values of $1-15 \times 10^{3} \  \mu$m$^3$\,s$^{-1}$ (for the mice data) agree with the only paper found to contain suitable experimental mice flow rates, which reported $2.22 \pm 1.985 \times 10^{3} \  \mu$m$^3$\,s$^{-1}$~\cite{Boster23flow}. Some murine model studies support this range (e.g.~\cite{Gan23}), and others argue values a number of magnitudes smaller (e.g.~\cite{Asgari16}).

\newpage
%
%
%\bibliography{csf_references}

\end{document}